\documentclass[12pt,preprint]{aastex}

\begin{document}

\def\hal{{H\alpha} }
\def\av{A_V}
\def\lum{{\mathcal{L}}_{bol}}
\def\logten{log_{10}}
\def\na{\ion{Na}{1}\ }
\def\pot{\ion{K}{1}\ }
\def\kms{km s$^{-1}$\ }
\def\kmsp{km s$^{-1}$ pix$^{-1}$\ }
\def\cms{cm s$^{-1}$\ }
\def\cmc{cm$^{-3}$\ }
\def\cmss{cm$^{2}$ s$^{-1}$\ }
\def\cmcs{cm$^{3}$ s$^{-1}$\ }
\def\msun{M$_\odot$\ }
\def\lsun{L$_\odot$\ }
\def\mj{M$_J$\ }
\def\teff{T$_{e\! f\! f}$~}
\def\gv{{\it g}~}
\def\vsini{{\it v}~sin{\it i}~}
\def\vrad{v$_{\it rad}$~}
\def\lbol{L_{\it bol}}
\def\lhal{L_{H\alpha}}
\def\fhal{F_{H\alpha}}
\def\fbol{F_{\it bol}}
\def\eqwhal{EW_{H\alpha}}
\def\h2o{H$_2$O}

\title{Magellan Echelle Spectroscopy of TW Hydrae Brown Dwarfs}

\author{Subhanjoy Mohanty}
\affil{Harvard-Smithsonian Center for Astrophysics, Cambridge, MA 02138, USA.}
\author{Ray Jayawardhana}
\affil{Department of Astronomy, University of Michigan, Ann Arbor, MI 48109, USA.}
\and
\author{David Barrado y Navascu\'{e}s}
\affil{Laboratorio de Astof\'{\i}sica Espacial y F\'{\i}sica Fundamental, INTA, 28080 Madrid, Spain.}

\begin{abstract} 
We present high-resolution optical spectroscopy of four candidate
members of the nearby TW Hydrae young association including three
brown dwarfs (2MASS 1207-3932, 2MASS 1139-3159 and TWA 5B) and one T
Tauri multiple star (TWA 5A). Using echelle spectra from the
Magellan Baade 6.5-meter telescope, we confirm the pre-main sequence
status and cluster membership of the substellar candidates, through
the detection of LiI, NaI consistent with low gravity, and radial 
velocity.  Given their late spectral type
($\sim$M8) and the youth of the association (age $\sim$ 10Myr), 
cluster membership certifies these three objects as very low-mass 
young brown dwarfs.  One of them (2MASS 1207-3932) shows
strong emission both in the Hydrogen Balmer series (H$\alpha$ to
H$\epsilon$) and in HeI (4471, 5876, 6678 and 7065\AA), compared to
other young brown dwarfs of similar spectral type.  The H$\alpha$ line
is also relatively broad (10\% width $\sim$ 200 \kms) and asymmetric.
These characteristics suggest that 2MASS 1207-3932 is a (weak)
accretor.  While we cannot rule out activity, comparison to a flaring 
field dwarf implies that such activity would have to be quite anomalous.
The verification of accretion would make 2MASS 1207-3932 the oldest 
actively accreting brown dwarf known to date, suggesting that inner 
disk lifetimes in substellar objects can be comparable to those in 
stars, consistent with a similar formation mechanism.  Finally, 
TWA 5A appears to be a variable accretor: observations
separated by two days show broad accretion-like H$\alpha$
(10\% width $\sim$270 \kms), with significant changes 
in the H$\alpha$ profile, as well as in the strengths of HeI, Na D and 
[OI6300].  TWA 5A is known to be a close triple; thus, 
our result implies that long-lived disks can exist even in multiple systems.  
\end{abstract}


\section{Introduction}
The recent identification of several groups of young stars within 
100 parsecs of the Sun has generated widespread interest (Jayawardhana
\& Greene 2001). Given their proximity and age differences, these groups 
are ideally suited for detailed studies of the origin and 
early evolution of stars, brown dwarfs (BDs) and planets. 
Perhaps the most intensely studied among these groups is the TW Hydrae 
Association (TWA), which consists of $\sim$20 co-moving stars 
(Zuckerman et al. 2001) at a distance of 47--67 pc 
and dispersed over some 20 degrees on the sky. The members are mostly 
late-type (K and M) stars, and include several 
interesting multiple systems (Brandeker, Jayawardhana \& Najita 2003, 
BJN03 hereafter, and references therein) and one A star. At an age 
of $\sim$10 Myr, the TWA fills a significant gap in the age sequence 
between $\sim$1-Myr-old T Tauri stars in molecular clouds like 
Taurus-Auriga and the $\sim$50-Myr-old open clusters 
such as IC 2391. That is particularly useful for deriving 
strong constraints on disk evolution timescales. Their diverse disk 
properties suggest that the TWA stars are at an age when disks 
are rapidly evolving, through coagulation of dust and dissipation of 
gas (Jayawardhana et al. 1999a; 1999b).  

Lowrance et al. (1999) found a BD candidate $\sim$2\arcsec~ 
from TWA 5A (CD -33$^{\circ}$ 7795); TWA 5A and 5B are now confirmed 
as a common proper motion pair (BJN03 and 
references therein). Recently, Gizis (2002) reported two isolated 
substellar candidates from the 2-Micron All-Sky Survey that may be 
members of the TWA. Together, these three objects
constitute a unique sample to explore the evolution 
of BD characteristics on a 10-Myr timescale.
%
%
Here we report high-resolution optical spectroscopy that confirm 
the youth, group membership, and substellar status of these three
objects. We also present a spectrum of the T Tauri multiple system 
TWA 5A, to which TWA 5B is bound. We use these spectra 
to investigate accretion, rotation and chromospheric activity.

\section{Observations \& Analysis}
We obtained high resolution optical spectra  using the Magellan
 Inamori Kyocera Echelle spectrograph (Bernstein et al. 2002) on the Baade
 6.5-meter telescope at Las Campanas Observatory, Chile in 2003 May. 
 Consecutive spectra were obtained for the three BD candidate members: 
3$\times$1200s for 2MASS 1207-3932 (May 8), 3$\times$1500s for 
2MASS 1139-2649 (May 9), and 2$\times$1800s for TWA 5B (May 10). 
 Additionally, we obtained two 600s exposures of TWA 5A, one each on
 May 8 and 10.  The spectra of the 2MASS objects (from now on, 2M1207 and 2M1139)
and the May 8 spectrum of
 TWA 5A were taken with a 1\arcsec-wide by 5\arcsec-long slit.  
The May 10 spectra of TWA 5A and 5B were obtained with a narrower 
0.7\arcsec$\times$5\arcsec~ slit.  The separation between TWA 5A 
and 5B is $\sim$ 2\arcsec, and TWA 5B is $\sim$ 7 magnitudes fainter 
in the optical.  To ensure no contamination of the 5B spectrum by 5A, 
therefore, we observed 5B with a narrower slit (0.7\arcsec),
 under optimal seeing conditions (better than 0.5\arcsec), with the 
slit positioned roughly perpendicular to the 5A-5B axis.  
%
%
The coverage obtained was $\sim$ 3200--4800\AA~ in the blue, and $\sim$ 
4800--8800\AA~ in the red, with overlapping orders.  The spectra are 
unbinned in wavelength, and binned by 2 pixels in the spatial direction.  
The 1\arcsec~ slit yielded a spectral resolution of $R\sim$19000 in the 
red, and 25000 in the blue; the 0.7\arcsec~ slit resulted in $R\sim$27000 
and 36000 respectively.  The data were reduced in standard fashion using IDL 
routines.  We derive rotational velocities (\vsini) for our 
targets by cross-correlating with a `spun-up' template of a slowly rotating 
standard.  We used a combination of dwarf and giant spectra for the 
template (see Mohanty \& Basri 2003).  Radial velocities (\vrad) were 
found by cross-correlating against the M6 dwarf Gl 406 
(\vrad$\approx$19 \kms). The \vsini and \vrad are listed in Table 1.

\section{Results \& Discussion}

\subsection{Membership \& Substellar Status}
All three TWA BD candidates have spectral types of $\sim$M8-M8.5 (Gizis 2002; Webb
et al. 1999), consistent with the features in our high-resolution data (e.g., 
strong TiO bands).  In all three, we detect LiI 6708 \AA. They also exhibit narrow 
NaI ($\sim$ 8200\AA) absorption profiles indicative of low gravity (intermediate 
between giants and dwarfs), and strong dMe-like $\hal$ emission (Fig. 1; 
equivalent widths --EW-- for $\hal$ and LiI given in Table 1).  In concert, these 
facts confirm the Pre-Main Sequence (PMS) status of TWA 5B, 2M1207 
and 2M1139: we can confidently exclude field M dwarfs (no LiI), low-mass field BDs  
with undepleted Li (type $\sim$ L2 or later; gravity $\gtrsim$ dwarfs values), 
Li-rich giants (much lower gravity; $\hal$ in absorption) and sub-giants (MS lifetime of 
M dwarfs too long for M-type subgiants to have formed yet).
 
For all three objects, our derived \vrad are commensurate with those found
 for other, bona-fide members of the association (Torres et al. 2003). 
TWA 5B is also confirmed as a common proper motion companion to TWA 5A 
(BJN03 and references therein). Gizis (2002) has found a proper motion 
consistent with membership for 2M1207.  In conjunction with our 
\vrad measurements and confirmation of PMS status, these observations 
verify TWA membership for both TWA 5B and 2M1207.  For 2M1139,
 Gizis (2002) found a proper motion apparently inconsistent
 with membership.  However, he notes that the astrometric uncertainty is
 large.  Therefore, given its PMS status and a \vrad consistent with 
other members, we consider it likely that 2M1139 also belongs to the
 TWA.  More precise astrometric measurents are required to verify this.  

All objects with detected Lithium and \teff less than about 2800K are
 expected to be substellar, regardless of age (see Basri 2000).
  The M8-8.5 spectral type of 2M1207, 2M1139 and TWA 5B, which implies 
\teff $<$ 2800K 
(e.g., Luhman 1999), therefore ensures that they are BDs, since they all show 
Li.  An age is required, however, to derive a mass.
  Since we confirm association membership for at least 2M1207 
and TWA 5B, an age of $\sim$ 10 Myr is justified for them. 
Comparing to the evolutionary tracks of Chabrier et al. (2000) then  yields 
$\sim$ 35 
Jupiters for both.  The same value is obtained for 2M1139, if it too is a member.  Even 
if it is not, one can still use the fact of Li detection, and a \teff estimate 
($\sim$2700K; Luhman 1999), to put upper limits on mass and age of $\sim$ 65 Jupiters and 
$\sim$ 250 Myr respectively, using the same models.  

\subsection{Accretion}
At very low masses and weak accretion rates, usual accretion indicators
(e.g., Stark broadening of $\hal$ wings, emission in OI and CaII) may 
disappear.  Simultaneously, other lines commonly associated with 
accretion at higher masses (e.g., HeI, upper Balmer lines) may appear 
in non-accreting, chromospherically active low-mass objects due to 
the lowering of the photospheric continuum with decreasing $T_{eff}$ (e.g., 
Jayawardhana, Mohanty, \& Basri 2003, hereafter JMB03). Consequently, 
it is difficult to distinguish between accretion and activity in weakly 
accreting low-mass objects.

Here we conservatively identify probable accretors based on asymmetric 
and broad $\hal$ (10\% width $\gtrsim$200\kms, following JMB03) 
{\it and} the detection of usual indicators of accretion
at larger masses, such as HeI and upper Balmer lines, at a level 
{\it higher} than the average for a given spectral type
(the same criteria adopted by Muzerolle et al. 2003). 
 By these conditions, 2M1139 and TWA 5B are chromospherically 
active, but not accreting.  2M1207 and TWA 5A, however, do appear
 to be accretors, as discussed below.  

{\bf{\it 2MASS 1207-3932}:}  As Fig. 2 shows, significant emission is seen
in HeI and in the upper Balmer series lines upto H$\epsilon$ in this object.
A perusal of the recent study by Muzerolle et al. (2003), meanwhile, shows 
that non-accreting M7-8.5 objects in their sample usually do not show HeI (at 
6678\AA, the only HeI line they include), or Balmer lines beyond 
$H\beta$.  Next, in Fig. 3, we show the normalized $\hal$ profiles 
from our 3 consecutive spectra of 2M1207.
 A clear asymmetry is seen in the first and third profiles, 
with much enhanced emission blueward of line-center.  In
 the second profile, the full-width at 10\% of peak flux is $\sim$ 200
 \kms: not unreasonable for very low-mass, low accretion-rate objects,
 and similar to that seen by JMB03 in a couple of mid- to late M accretors. 
 In the first and last profiles, the width is slightly less,
 at $\sim$ 170 \kms.  However, superimposing the three spectra (not shown) 
reveals that the line wings are exactly the same; the smaller 
10\% width in the first and last is artificially induced simply
 by the increase in peak flux.  We therefore adopt $\sim$ 200\kms
 as the more robust value. By all our criteria, therefore, 
2M1207 appears to be (weakly) accreting.
Can the emission in this object simply be a result of enhanced
activity compared to other, similar spectral type objects?  While we
cannot rule this out, the following argument makes this seem unlikely.
In one of our high-resolution spectra of the field M7.5 dwarf LHS 2397A, 
taken during a flare,  the $\hal$ EW is $\sim$ 30 \AA~ (Mohanty 
\& Basri 2003), i.e., almost exactly the same as observed in 2M1207.  
If the emission in 2M1207 were due to chromospheric activity, we would 
expect its line profiles to look very similar to those of LHS 2397A. 
given the nearly identical spectral types. 
The line profile comparison is shown in Fig.3.  Clearly, the LHS 2397A 
$\hal$ profile much more symmetric compared to the first and last 2M1207 
spectra, and significantly narrower in all three cases: 2M1207 exhibits
broader $\hal$ line wings, as expected from accretion.
The HeI line at 6678\AA~ is also much stronger in 2M1207 than in
LHS 2397A; since the spectral types (and hence the underlying
continuum) are almost the same, the actual flux in HeI emission is
thus also higher.  This too indicates that the same physical process 
is not responsible for the emission in the two objects; in particular, 
excess HeI emission can arise from the very hot accretion-shock region.  
We suggest, therefore, that 2M1207 is a bona-fide accretor. 
However, our arguments against activity are not ironclad, and the presence
 of accretion needs to be checked through other diagnostics. 
The lack of a $K-L^\prime$ excess in 2M1207 (Jayawardhana et al. 2003) 
could be the result of a nearly edge-on disk inclination or grain growth. 

If accretion in 2M1207 is confirmed, it would suggest that inner disk 
lifetimes of substellar objects can be comparable to those of their 
stellar counterparts, and further strengthen the case for a common
formation mechanism for BDs and stars.   

{\bf{\it TWA 5A}:}  The 10\% full-width of the $\hal$ line in TWA 5A
 is $\sim$ 270 \kms, which is commensurate with accretion (e.g., JMB03).
 However, this object is known to be a triple, with a $\sim$ 0''.06 
binary resolved by adaptive optics (BJN03) as well as a spectroscopic
 companion (Torres et al. 2003). 
To discount the possibility that 
the $\hal$ line is broadened to accretion-like widths simply due to 
blending of $\hal$ from the three stars, we compare our two spectra of 
TWA 5A, obtained two days apart.  Fig. 4 shows that the spectra are 
exactly the same, {\it except} in lines that are accretion indicators. 
 This implies that the spectral change is not due to variations in 
line-blending, but in the intrinsic parameters of (at least) one of 
the TWA 5A components.  In particular, we see that the first spectrum 
(in black) shows excess red-shifted emission and blue-shifted absorption 
in $\hal$, HeI5876\AA{ } and Na D, as well as excess emission in [OI6300],
 compared to the second spectrum (in grey).  Though $\hal$, NaD and
 HeI emission may conceivably arise from chromospheric activity, the 
blue-shifted absorption seen in these three lines strongly suggests 
accretion.  Furthermore, [OI6300] is an 
excellent diagnostic of outflowing winds associated with accretion; 
it is often seen in accreting CTTs, but never in non-accreting WTTs 
(Muzerolle et al. 2003).  All this supports ongoing accretion in the 
TWA 5A system.  The fact that the two spectra differ significantly in
 the accretion diagnostics implies that the process is variable.  Our results also 
suggest that inner disks can be long-lived even in close multiple systems.

\subsection{Rotation \& Activity}
In field dwarfs down to $\sim$ M8, a saturation-type rotation-activity
connection is seen: those rotating faster than $\sim$ 5 \kms exhibit
saturated levels of chromospheric $\hal$ emission, while slower
rotators show a range of smaller $\hal$ fluxes (Mohanty \& Basri
2003).  For our young sample, we will discuss this
issue in more detail in a future paper, after deriving accurate \teff in order to 
convert the observed $\hal$ EWs to fluxes.  However, we note here that the $\hal$ EW 
observed in
our three M8 PMS objects are, at the very least, comparable to those
seen in saturated field dwarfs of the same spectral type; this is in
keeping with their moderately rapid rotation ($>$ 10\kms).  In 2M1207, 
the $\hal$ emission is far stronger than in merely
saturated field dwarfs; as discussed above, the EW in
this case is comparable to that in strongly flaring M8 dwarfs.
Similar EWs have been seen in some other young, (apparently)
non-accreting late-M objects, lending support to the idea that
activity may be enhanced in young low-mass objects compared to field
objects of similar spectral type (e.g.,
Jayawardhana, Mohanty, \& Basri 2002, 2003).  However, the
$\hal$ profile in these cases is narrow and symmetric.  If 2M1207 
were a non-accretor, then the strong asymmetries seen in at least two of 
its $\hal$ profiles would be puzzling.  

\acknowledgments
We are most grateful to Gibor Basri for insightful discussions. We thank
the referee Jeff Valenti for a prompt review and the Magellan staff 
for their assistance. S.M. is indebted to the SIM-YSO group for the 
postdoctoral fellowship that made this research possible.  This work was 
supported in part by NSF grant AST-0205130 and NASA grant NAG5-11905 to 
R.J.

\clearpage

\begin{figure}
\epsscale{0.55}
\plotone{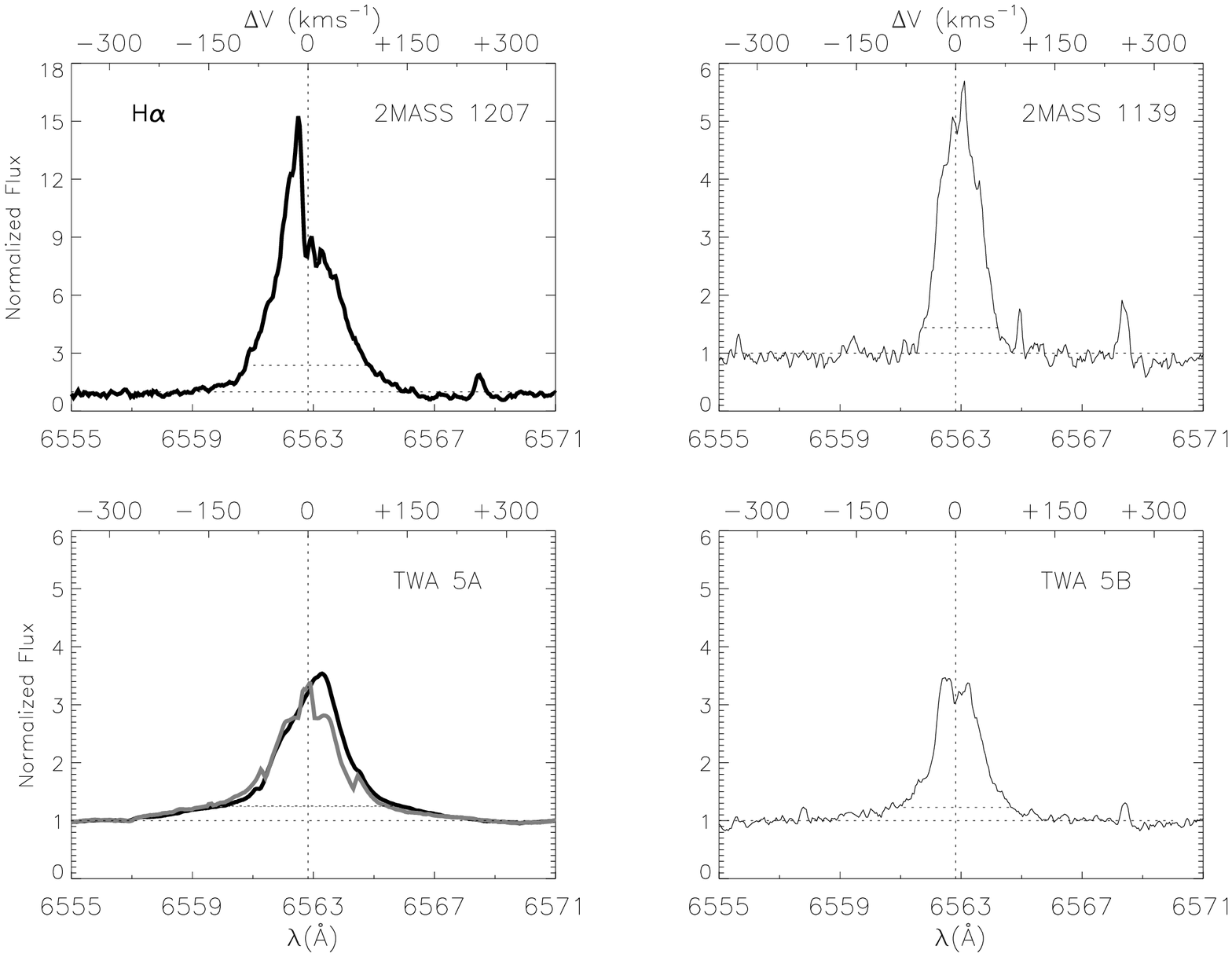}
\caption{Averaged $\hal$ profiles for our targets (both profiles shown for TWA 5A).
  Thick lines indicate the two probable accretors.  The horizontal dashed lines
 indicate the continuum and 10\% of peak emission.  All profiles are shifted to
 zero-velocity (approximate for TWA 5A; we have not calculated its \vrad).}  
\end{figure}

\begin{figure}
\epsscale{0.70}
\plotone{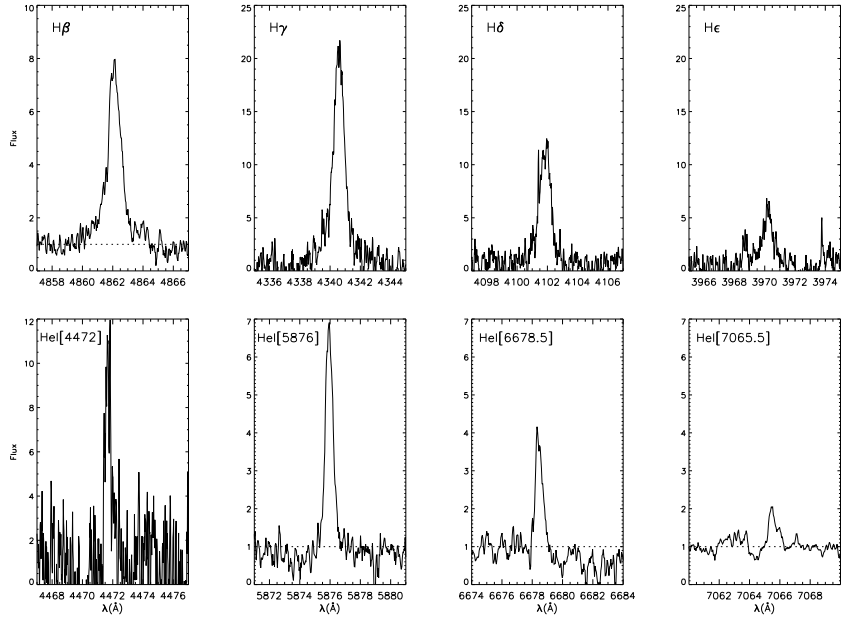}
\caption{Averaged HI and HeI emission in 2MASS 1207.  A continuum was 
detected around only some of these lines; these have been continuum-normalized 
(continuum: dashed horizontal line).  In the others, no true continuum was 
detected, making flux or EW measurements impossible; these have only been 
divided by the r.m.s. of the surrounding noise for clarity. 
%
%
}
\end{figure}

\clearpage

\begin{figure}
\epsscale{0.5}
\plotone{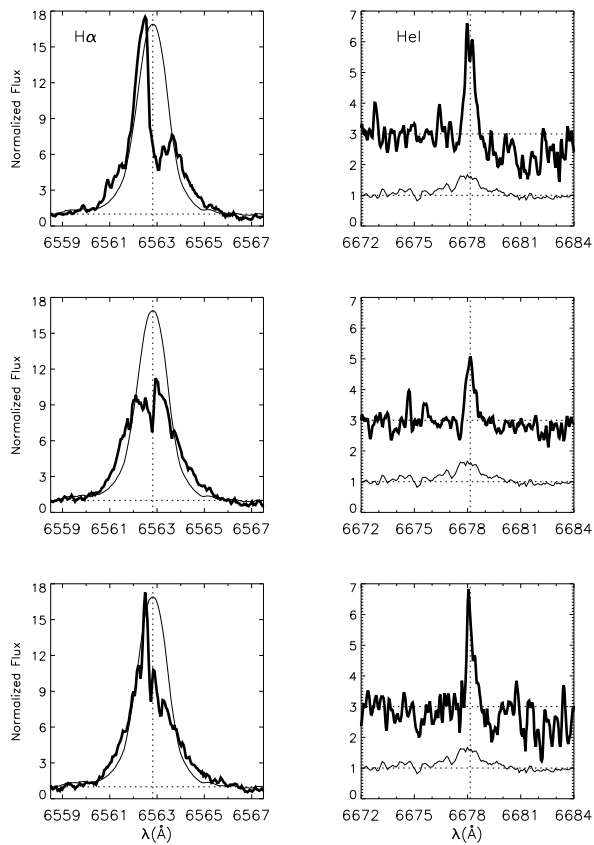}
\caption{{\it Top to bottom, left panels}:  $\hal$ from our 3 spectra
of 2MASS 1207 (thick lines) compared to the $\hal$ in the flaring
field M7.5 dwarf LHS 2397A (observed at similar resolution on Keck HIRES). 
%
%
{\it Top to bottom, right panels}:  HeI6678\AA{} comparison.
%
}
\end{figure}

\begin{figure}
\epsscale{0.50}
\plotone{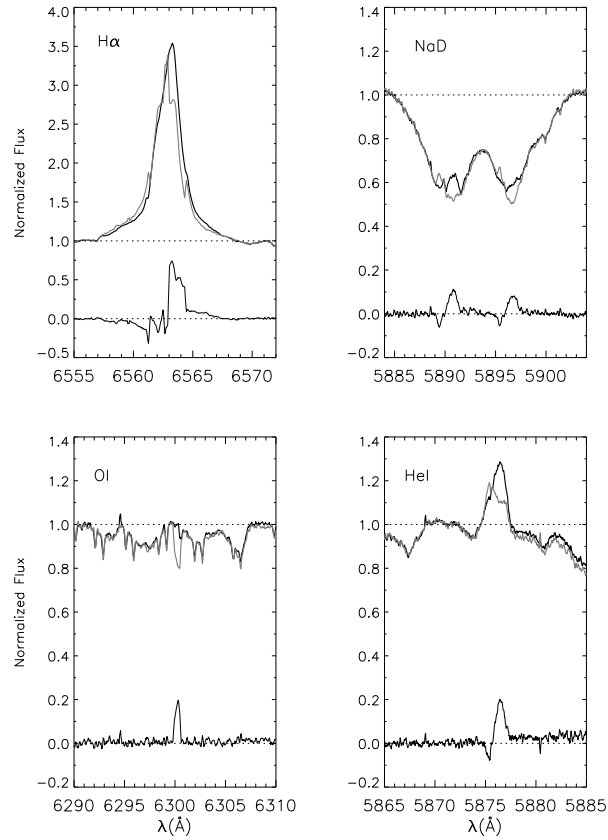}
\caption{Comparison of two TWA 5A spectra, in regions containing
accretion-indicating lines ($\hal$, NaD, [OI6300], HeI5876).  
The two spectra are exactly the same, except one of them 
(black) shows excess emission in $\hal$, NaD, OI \& HeI, and 
blue-shifted absorption in $\hal$, NaD \& HeI, compared to the other 
spectrum (grey).}
\end{figure}
\clearpage

\bigskip
\begin{center}
\begin{deluxetable}{lcccccc}
\tablecaption{\label{tab1} Spectroscopic Properties of TWA Targets}
\tablehead{
\colhead{Object} &
\colhead{Sp Type\tablenotemark{a}} &
\colhead{\vrad} &
\colhead{\vsini} &
\colhead{LiI EW\tablenotemark{b}} &
\colhead{H$\alpha$ EW\tablenotemark{b}} &
\colhead{H$\alpha$ 10\% FW\tablenotemark{c}} \\
 & & (\kms) & (\kms) & (\AA) & (\AA) & (\kms) \\}

\startdata
2MASS 1207-3932  & M8  & 11.2$\pm$2 & 13$\pm$2 & 0.5 & 27.7 & 204$\pm$10 \\
2MASS 1139-3159  & M8  & 11.6$\pm$2 & 25$\pm$2 & 0.5 & 7.3  & 111$\pm$10 \\
TWA 5B     &$\sim$M8.5 & 13.4$\pm$2 & 16$\pm$2 & 0.3 & 5.1  & 162$\pm$10 \\
TWA 5A\tablenotemark{d} & M1.5  & -- & --   & 0.6 & 7.2/6.6\tablenotemark{e}  & 266$\pm$10 \\
\enddata
\tablenotetext{a}{Spectral types for 2MASS objects from Gizis (2002) and for TWA 5A \& 5B from Webb et al. (1999).} 
\tablenotetext{b}{{\it Pseudo}-equivalent widths, for LiI \& $\hal$; estimated $\hal$ EW errors $\lesssim$ 10\%.}
\tablenotetext{c}{For 2M1139 \& TWA 5B, we quote the average width from
 their consecutive spectra over a single night.  For 2M1207,
 we adopt the $\hal$ 10\% width of only the second spectrum as the 
realistic value (see text); the average width from 
all three spectra is 187$\pm$10 \kms.  For TWA 5A, we quote 10\% FW 
values for both nights; they are very similar.} 
\tablenotetext{d}{We did not derive \vrad or \vsini for the TWA 5A system, given 
its known \vrad variability (Torres et al. 2003), and the questionable accuracy of 
\vsini given possible line blending.}
\tablenotetext{e}{In TWA 5A, the larger $\hal$ EW is
 for the spectrum with excess accretion-related emission. }  
\end{deluxetable}
\end{center}

\end{document}